\documentstyle[12pt]{article}

\advance\hoffset by -1.0cm
\def\3{\ss}

\def\sq{\hbox{\rlap{$\sqcap$}$\sqcup$}}
\def\qed{\ifmmode\sq\else{\unskip\nobreak\hfil
\penalty50\hskip1em\null\nobreak\hfil\sq
\parfillskip=0pt\finalhyphendemerits=0\endgraf}\fi}

\def\bbbc{{\mathchoice {\setbox0=\hbox{$\displaystyle\rm C$}\hbox{\hbox
to0pt{\kern0.4\wd0\vrule height0.9\ht0\hss}\box0}}
{\setbox0=\hbox{$\textstyle\rm C$}\hbox{\hbox
to0pt{\kern0.4\wd0\vrule height0.9\ht0\hss}\box0}}
{\setbox0=\hbox{$\scriptstyle\rm C$}\hbox{\hbox
to0pt{\kern0.4\wd0\vrule height0.9\ht0\hss}\box0}}
{\setbox0=\hbox{$\scriptscriptstyle\rm C$}\hbox{\hbox
to0pt{\kern0.4\wd0\vrule height0.9\ht0\hss}\box0}}}}

\begin{document}
\thispagestyle{empty}
\begin{flushright}
DAMTP-94-4 \\
hep-th/9401166
\end{flushright}
\begin{center}
\vspace{2.5cm}

{\huge A general transformation formula for conformal fields}
\vspace{1.0cm}

{\large Matthias Gaberdiel}
\footnote{e-mail: M.R.Gaberdiel@amtp.cam.ac.uk} \\
{Department of Applied Mathematics and Theoretical
Physics\\
University of Cambridge, Silver Street \\
Cambridge, CB3 9EW, U.\ K.\ }
\vspace{1.0cm}

{January 1994}
\vspace{3.5cm}

{Abstract}
\end{center}

An explicit transformation formula for chiral conformal fields
under arbitrary holomorphic coordinate transformations
is established.
As an application I calculate the transformation
law of the general quasiprimary field at level $4$.
\newpage

\noindent {\bf 1. Introduction}
\bigskip

An important feature of two-dimensional conformal field theory
is the fact that the algebra of conformal transformations
can be understood as the direct product of two infinite
dimensional algebras. The theory factorises therefore
into a holomorphic and an anti-holomorphic subtheory,
which have the group of analytic substitutions of the
(anti-)analytic variable $z (\bar{z})$ as reparametrisation
symmetries \cite{BPZ}. The central extension of this
algebra is the Virasoro algebra $L_{n}$, which is
thus the remains of the conformal symmetry
for each of the chiral theories.

The Virasoro algebra contains the algebra of M\"{o}bius
transformations $L_{j}, j=-1,0,1$, namely the algebra
of automorphisms of the Riemann sphere, as a subalgebra.
For the M\"{o}bius subgroup the transformation formula of (chiral)
conformal fields is known \cite{Peter89}. Namely, for
$\psi$ a $su(1,1)$ highest weight vector of conformal
weight $h$, i.~e.\
$L_{1}\;\psi=0$ and $L_{0}\;\psi=h\; \psi$, the
corresponding vertex operator transforms as
\begin{equation}
\label{Mob}
D_{\gamma} V(\psi,z) D_{\gamma}^{-1}=
\left[ \frac{d\gamma(z)}{dz}\right]^{h} V(\psi,\gamma(z)),
\end{equation}
where
\begin{equation}
\label{Mob1}
D_{\gamma}=exp\left\{\frac{b}{d} L_{-1}\right\}
\left(\frac{\sqrt{ad-bc}}{d}\right)^{2 L_{0}} exp\left\{-\frac{c}{d}
L_{1}\right\}
\end{equation}
and the M\"{o}bius transformation is given by
\begin{equation}
\gamma(z) = \frac{az + b}{cz+ d}.
\end{equation}
\smallskip

In this letter I want to show how arbitrary holomorphic
transformations can be implemented in the operator
framework of (chiral) conformal field theory.
\vspace*{1.0cm}

\noindent {\bf 2. The main result}
\bigskip

Let $\psi$ be a primary field of weight $h$, i.\ e.\
$L_{0}\; \psi=h\; \psi$ and $L_{n}\; \psi=0$ for $n>0$. Let
$f$ be an analytic function defined in some open neighbourhood
of the origin $f:D\rightarrow \bbbc$ which is locally invertible, i.\ e.\
$f'(0)\neq 0$. Let furthermore the action of the exponentials
of the Virasoro generators be local with respect to the given
system of vertex operators in the sense of \cite{Peter89}.
Then the local transformation of the primary field $\psi$ is given
as
\begin{equation}
\label{Theorem1}
e^{f(0) L_{-1}}\; f'(0)^{L_{0}} \prod_{n=1}^{\infty} e^{S_{n} L_{n}}
V(\psi,z) \prod_{n=\infty}^{1} e^{-S_{n} L_{n}}\; f'(0)^{- L_{0}}\;
e^{- f(0) L_{-1}}  =
f'(z)^{h} \; V(\psi,f(z)).
\end{equation}
The $S_{n}$ are defined to be $S_{n}=T_{n}(z)\left|_{z=0}\right.$
and the $T_{n}(z)$ are recursively defined
\begin{equation}
\label{T0}
T_{0}(z)=\log{f'(z)} ,
\end{equation}
$$ T_{n}(z)=\frac{1}{n+1} \left( T_{n-1}'(z) - A_{n}(z)
\right)\hspace{2.0cm} n\geq 1. $$
$A_{n}(z)$ is of the form
\begin{equation}
\label{An}
A_{n}(z)=\sum_{\sum_{i=1}^{m} k_{i} l_{i}=n} C^{m}(k_{1},l_{1};\ldots;
k_{m},l_{m})\; T_{k_{1}}^{l_{1}}(z) \cdots T_{k_{m}}^{l_{m}}(z),
\end{equation}
where the sum extends over those $2m$-tupels, where
$1\leq k_{1}<k_{2}<\cdots < k_{m}$ and $l_{i}$ is positive. The coefficients
$C^{m}(k_{1},l_{1};\ldots;k_{m},l_{m})$ are zero unless
$K_{p}>k_{p+1}$ for
$p=1, \ldots , m-1$, where
\begin{equation}
K_{p}:=\Bigl( \sum_{i=1}^{p} k_{i} l_{i} \Bigr) -1.
\end{equation}
If this condition is met they are given as
\begin{equation}
\label{C1}
C^{1}(k,l)=\left\{
\begin{array}{cl}
{\displaystyle \frac{k+1}{l\, (l-2)! }\;\; \prod_{p=1}^{l-2} (pk-1)}
& l\geq 3
\vspace*{0.2cm} \\

{\displaystyle \frac{k+1}{2}} & l=2
\vspace*{0.2cm} \\

{\displaystyle 0} & l=1,
\end{array}
\right.
\end{equation}
and for $m\geq 2$ as
\begin{equation}
\label{Cn}
C^{m}(k_{1},l_{1};\ldots ;k_{m},l_{m})=
\left( \prod_{j=2}^{m} \; \frac{1}{l_{j}!}\;  \prod_{q_{j}=1}^{l_{j}}
\Bigl(K_{j-1} + (q_{j}-2) k_{j} \Bigr) \right) C^{1}(k_{1},l_{1}).
\end{equation}
\medskip

\noindent Before proving the formula I would like to remark that
$3!\, T_{2}(z)$ is the Schwarzian derivative of the function $f$
\begin{equation}
(Df)(z) = \frac{f'''(z)}{f'(z)} - \frac{3}{2} \left( \frac{f''(z)}
{f'(z)} \right)^{2}.
\end{equation}
Because of (\ref{C1}), $C^{1}(1,l)=0$ for $l\neq 2$ and
$C^{m}(1,2; \ldots)=0$ for $m\geq 2$, as $K_{1}=1 \leq k_{2}$.
Thus all $S_{n}, n\geq 2$ are
sums of products of derivatives of the Schwarzian derivative.
This implies that in the case of the M\"{o}bius transformations
the formula (\ref{Theorem1}) reduces to (\ref{Mob}),
as the Schwarzian derivative
vanishes if and only if $f$ is a M\"{o}bius transformation.
\medskip

To prove the formula it is enough to consider the case
where $f(0)=0$ and $f'(0)=1$, as I can use
(\ref{Mob},\,\ref{Mob1}).
By the uniqueness theorem of \cite{Peter89} it is
furthermore sufficient to check the identity
when applying both sides to the vacuum.

The strategy of the proof is to show (by induction)
that all coefficients of the power series expansions in
$z$ agree. The induction start is trivial, since both sides
are just $\psi$. To prove the induction step suppose that
the coefficients for the power series expansion agree
up to order $n-1$. Taking the
$n$-th derivative on the left-hand-side, evaluated at
zero, I get
\begin{equation}
e^{S_{0} L_{0}}  \cdots e^{S_{n} L_{n}} (L_{-1})^{n} \psi,
\end{equation}
which equals
\begin{eqnarray}
\label{10}
{\displaystyle L_{-1} \left\{ e^{S_{0} L_{0}}  \cdots
   e^{S_{n-1} L_{n-1}} (L_{-1})^{n-1} \psi\right\}} \hspace*{7cm} \\
\label{11}
{\displaystyle + \sum_{k=0}^{n} e^{S_{0} L_{0}} \cdots
e^{S_{k-1} L_{k-1}} [e^{S_{k} L_{k}},L_{-1}] e^{S_{k+1} L_{k+1}}\cdots
e^{S_{n-1} L_{n-1}} (L_{-1})^{n-1} \psi.}
\end{eqnarray}
Using the formula
\begin{equation}
\label{12}
[e^{A},B]=\int_{0}^{1} dt\;  e^{(1-t)\,A}\; [A,B]\; e^{t\,A}
\end{equation}
I can rewrite this expression as
\begin{eqnarray}
{\displaystyle L_{-1} \left\{ e^{S_{0} L_{0}} \cdots
   e^{S_{n-1} L_{n-1}} (L_{-1})^{n-1} \psi\right\}}
\hspace*{8cm} \nonumber \\
\label{13}
{\displaystyle
+ \sum_{k=1}^{n} (k+1)\, S_{k}\, e^{S_{0} L_{0}} \cdots e^{S_{k-1} L_{k-1}}
L_{k-1} e^{S_{k} L_{k}} \cdots e^{S_{n-1} L_{n-1}} (L_{-1})^{n-1}\psi} \\
\label{14}
{\displaystyle
+ \sum_{k=1}^{n} (k+1)\, S_{k} \int_{0}^{1} dt  \;
e^{S_{0} L_{0}} \cdots e^{S_{k-1} L_{k-1}}
[e^{(1-t)S_{k} L_{k}},L_{k-1}]} \hspace*{4cm} \nonumber \\
\hspace*{\fill} {\displaystyle
e^{t S_{k} L_{k}}
e^{S_{k+1} L_{k+1}} \cdots e^{S_{n-1} L_{n-1}} (L_{-1})^{n-1} \psi,}
\end{eqnarray}
since the $k=0$-term does not contribute because of $S_{0}=0$.
Using the definition of $S_{k}$ this expression now becomes
\newpage

\begin{eqnarray}
\label{Der1}
{\displaystyle L_{-1} \left\{ e^{S_{0} L_{0}} \cdots
e^{S_{n-1} L_{n-1}} (L_{-1})^{n-1} \psi\right\}}
\hspace*{8cm}  \\
\label{Der2}
{\displaystyle
+ \sum_{k=1}^{n} \left. T_{k-1}'(z) \right|_{z=0} \,
\frac{d}{d S_{k-1}}
\left\{ e^{S_{0} L_{0}} \cdots e^{S_{n-1} L_{n-1}} (L_{-1})^{n-1}\psi
\right\}} \hspace*{3cm} \\
\label{term1}
{\displaystyle
+ \sum_{k=1}^{n} (k+1)\, S_{k} \int_{0}^{1} dt\;
e^{S_{0} L_{0}} \cdots e^{S_{k-1} L_{k-1}}
[e^{(1-t)S_{k} L_{k}},L_{k-1}]} \hspace*{4cm} \\
\label{term2}
\hspace*{\fill} {\displaystyle
e^{t S_{k} L_{k}}
e^{S_{k+1} L_{k+1}} \cdots e^{S_{n-1} L_{n-1}} (L_{-1})^{n-1} \psi}
\\
{\displaystyle
- \sum_{k=1}^{n} A_{k}(0)\;
e^{S_{0} L_{0}} \cdots e^{S_{k-1} L_{k-1}}
L_{k-1} e^{S_{k} L_{k}} \cdots e^{S_{n-1} L_{n-1}}
(L_{-1})^{n-1}\psi.} \nonumber
\end{eqnarray}
(\ref{Der1}) and (\ref{Der2}) are just the $n$-th derivative of the
right-hand-side at $z=0$, since the term in brackets
is (by the induction hypothesis) the $(n-1)$st derivative
of the right-hand-side at $z=0$. Thus it remains to prove that
the remaining terms cancel. As the term $k=n$ in
(\ref{term1},\,\ref{term2}) does not contribute, $A_{n}$
has to solve
\begin{eqnarray}
\label{commu}
{\displaystyle
A_{n} L_{n-1} (L_{-1})^{n-1} \psi
= \sum_{k=1}^{n-1} (k+1)\, T_{k}^{2} \int_{0}^{1} dt^{(1)}\;
(1- t^{(1)})\; \int_{0}^{1} dt^{(2)} \;
e^{T_{0} L_{0}} \cdots e^{T_{k-1} L_{k-1}}} \\
\label{commu1}
{\displaystyle
e^{(1-t^{(1)}) (1 - t^{(2)}) T_{k} L_{k}} \; L_{2k-1} \;
e^{(1-(1-t^{(1)})(1-t^{(2)})) T_{k} L_{k}}
e^{T_{k+1} L_{k+1}} \cdots e^{T_{n-1} L_{n-1}} (L_{-1})^{n-1} \psi} \\
{\displaystyle
\label{16}
- \sum_{k=1}^{n-1} A_{k}\;
e^{T_{0} L_{0}} \cdots e^{T_{k-1} L_{k-1}}
L_{k-1} e^{T_{k} L_{k}} \cdots e^{T_{n-1} L_{n-1}}
(L_{-1})^{n-1}\psi,}
\end{eqnarray}
where I have used (\ref{12}) to rewrite the commutator and I have
replaced $S_{l}$ by $T_{l}$, since $A_{n}$ has to satisfy
the above equation for a neighbourhood of $z=0$.

To check (\ref{commu}-\ref{16}) I commute the Virasoro generator
$L_{2k-1}$ so that it stands next to the exponential containing
$L_{2k-1}$ to get a term of the form (\ref{16}). I thereby obtain
a sum of commutators of the form (\ref{term1}), which I can rewrite
using (\ref{12}) to get terms of the form
(\ref{commu},\,\ref{commu1}). I then
repeat the above procedure for each of these terms.
For a given $n$ only those terms contribute, whose
Virasoro generator $L_{p}$ satisfies $p\leq n-1$. I therefore
have to do the above algorithm only finitely many times.
Furthermore, in (\ref{commu}-\ref{16})
all terms with $p<n-1$ cancel, since
by induction $A_{p}$ with $p<n-1$ satisfies an equation
corresponding to (\ref{commu}-\ref{16}).
Hence there is a solution for $A_{n}$
and it has to be of the form (\ref{An}). To calculate the numerical
coefficients $C^{m}(k_{1},l_{1};\ldots;k_{m},l_{m})$ one
finally observes that the term containing
$T_{k_{1}}^{l_{1}} \cdots T_{k_{m}}^{l_{m}}$ can be obtained
in one way only, namely by successively commuting $L_{2k_{1}-1}$
with $e^{t' T_{k_{1}} L_{k_{1}}}$
($l_{1}-2$ times), then commuting the
resulting Virasoro generator  with $e^{T_{k_{2}} L_{k_{2}}}$
($l_{2}$ times) and so forth. Calculating the corresponding
multiple integrals (coming from the successive use of (\ref{12}))
I arrive at (\ref{C1}) and (\ref{Cn}).
\medskip

The proof of the formula holds formally
only for meromorphic fields, i.~e.\ fields with integral conformal
weight. However, as the action of the Virasoro
algebra is local with respect to arbitrary fields,
the result should extend to the more general case as well.
\vspace*{1.0cm}

\noindent {\bf 3. Consequences}
\bigskip

It was a priori to be expected that (\ref{Theorem1})
holds for some coefficients $S_{n}$. The interesting
new piece of information is therefore the explicit
formula for these coefficients (\ref{T0} - \ref{Cn}).
Furthermore, the above derivation suggests how the
result generalises to arbitrary conformal fields:
\medskip

\noindent Let $f: D\rightarrow \bbbc$ be a function as above and let the
$S_{n}$'s be defined as in (\ref{T0} - \ref{Cn}). Let $\phi$ be a
secondary field of weight $h+m$ in the family
of $\psi$. Let $H_{(p)}(f^{(i)}(z))$ be given as
\begin{equation}
\label{Lemma21}
f'(z)^{L_{0}} \prod_{n=1}^{m} e^{T_{n}(z) L_{n}} \; \phi
= \sum_{(p)} H_{(p)}(f^{(i)}(z)) \; L_{p_{1}}\cdots L_{p_{k}}\; \phi,
\end{equation}
where $p$ is a $k$-tupel of integers $p_{1}, \ldots\/, p_{k}$
and the sum extends over all $p$ for which
$L_{p_{1}}\cdots L_{p_{k}} \phi \neq 0$.
Then the transformation law for $\phi$ is given as
\begin{eqnarray}
\label{Lemma2}
{\displaystyle e^{f(0) L_{-1}} \;
f'(0)^{L_{0}} \prod_{n=1}^{\infty} e^{S_{n} L_{n}} \;
V(\phi,z) \; \prod_{n=\infty}^{1} e^{-S_{n} L_{n}}\;  f'(0)^{- L_{0}}\;
e^{- f(0) L_{-1}}}
\hspace*{2.5cm} \nonumber \\
\hspace*{5cm} {\displaystyle = \sum_{(p)} H_{(p)}(f^{(i)}(z)) \;
V(L_{p_{1}}\cdots L_{p_{k}} \phi,f(z)).}
\end{eqnarray}
\smallskip

The proof of (\ref{Lemma2}) is a replica of the above proof ---
the only difference being the fact
that to prove the induction step from $n-1$ to $n$
the sum in (\ref{11}) extends now to $k=n+m$ and
similarly in (\ref{commu}-\ref{16}). However, since the $A_{p}$ are defined
recursively, so as to cancel (identically) all terms of the
corresponding form, all remaining arguments go through.
\medskip

The above formulae show that the
product of exponentials in (\ref{Theorem1}),~resp.
(\ref{Lemma2}), implements arbitrary holomorphic
coordinate transformations in the operator framework
of chiral conformal field theory. They
provide thus a link between two
different concepts of primary/secondary/etc.\ fields: the one
defined in terms of transformation properties of fields
and the one defined in terms of annihilation properties
of the corresponding vectors under the action of the
Virasoro algebra.
\smallskip

Furthermore, (\ref{Lemma21},\,\ref{Lemma2}) is a powerful tool for the
determination of the transformation law for
arbitrary conformal fields. As an application I have
calculated the transformation formula for the
general quasiprimary field at level $4$ in the family
of the primary field $\psi$.
This fields is given as
\begin{equation}
\psi_{4}= \left( b_{4}\; L_{-1}^{4} + a_{4}\; L_{-2}^{2}
+ m_{1}\; L_{2}\; L_{-1}^{2} + m_{2}\; L_{-3}\; L_{-1}
+ m_{3}\; L_{-4} \right) \psi,
\end{equation}
where
\begin{eqnarray}
m_{1} & = &
{\displaystyle - 4 b_{4} \frac{2 h + 3}{3} } \\
m_{2} & = &
{\displaystyle \frac{4 b_{4} (4 h^2 + 8 h + 3) - 9 a_{4}}{6}} \\
m_{3} & = &
{\displaystyle \frac{- 4 b_{4} h  (4 h^2 + 8 h + 3)
                     + 9 a_{4} (h-1)}{15}}.
\end{eqnarray}

\noindent Using (\ref{Lemma21},\,\ref{Lemma2}) I find that
$\psi_{4}$ transforms as
\begin{eqnarray}
{\displaystyle
\psi_{4}(z)} & \rightarrow &
{\displaystyle f'(z)^{h+4} \; \psi_{4}(f(z))
+ \frac{a_{4}\; \beta_{1} + b_{4}\; \beta_{2}}
{a_{2}} \; (Df)(z)\; f'(z)^{h+2}\; \psi_{2}(f(z)) }
\nonumber \\
& &
{\displaystyle
+ \left(a_{4}\; \gamma_{1} + b_{4}\; \gamma_{2} \right)
\frac{2}{2h + 1}\; (Df)(z) \; f'(z)^{h+2} \; \psi''(f(z))} \nonumber \\
& &
{\displaystyle
+ \left(a_{4}\; \gamma_{1} + b_{4}\; \gamma_{2} \right)
\left( + 2\; (Df)(z) \; f''(z) \; f'(z)^{h}
- (Df)'(z)\; f'(z)^{h+1} \right) \psi'(f(z))} \nonumber \\
& &
{\displaystyle
+  \left[ \left(a_{4} \; \gamma_{1} + b_{4}\; \gamma_{2} \right) h
           \left( (Df)(z)\; f''(z)^{2}\; f'(z)^{h-2}
                    - (Df)'(z)\; f''(z)\; f'(z)^{h-1} \right.
\right. } \nonumber \\
& &
{\displaystyle \left. \left.
                    + \frac{1}{5}\; (Df)''(z) \;f'(z)^{h} \right)
+ (Df)^{2}(z) \left(a_{4} \; \delta_{1}
+ b_{4}\; h\; (\frac{4}{5}\; \gamma_{2} + \delta_{2}) \right)
\right]  \psi(f(z)),}
\end{eqnarray}
where $\psi_{2}$ is the general quasiprimary field at level $2$
\begin{equation}
\psi_{2}= a_{2}\; \left( L_{-1}^{2} -
\frac{2(2 h+1)}{3} L_{-2} \right)\psi
\end{equation}
and
\begin{eqnarray}
\beta_{1} & = &
{\displaystyle - \frac{5\; c + 58\; h + 22}{20\; (2 \;h + 1)}} \\
\beta_{2} & = &
{\displaystyle  \frac{8\; h^{3} + 36\; h^{2} + 36\; h}
{5\; (2 \; h + 1)}} \\
\gamma_{1} & = &
{\displaystyle \frac{1}{40} \left( 5\; c + 8\; h - 3 \right)} \\
\gamma_{2} & = &
{\displaystyle - \frac{1}{90} \left( 32\; h^3 - 76\; h^2 + 104\;  h
- 15  + 5\; c \left( 4\; h^2 + 8\; h + 3 \right) \right)} \\
\delta_{1} & = &
{\displaystyle \frac{c^2}{144} + \frac{c\; (40\; h + 11)}{360}
 + \frac{ h\; (544\; h - 329)}{900} } \\
\delta_{2} & = &
{\displaystyle \frac{ (8\; c\; h^{2} + 6\; c\; h - 9\; c
- 100\;  h^{2} + 75\; h)}{45}.}
\end{eqnarray}
In the case of the vacuum representation, i.~e.\ $\psi=\Omega$ and $h=0$,
$b_{4},\; m_{1}$ and $m_{2}$ can be chosen arbitrarily and
$m_{3}= - 3/5\; a_{4}$. The transformation formula
then simplifies to  \cite{RS88}
\begin{eqnarray}
\psi_{4}^{0}(z) & \rightarrow &
{\displaystyle f'(z)^{4}\; \psi_{4}^{0}(f(z)) + a_{4}\;
\frac{5\;c + 22}{30}\; (Df)(z)\; f'(z)^{2}\; T(f(z))} \nonumber \\
& &
{\displaystyle + \frac{c}{12} \; \frac{1}{2} \;
a_{4}\; \frac{5\;c + 22}{30}\; (Df)^{2}(z)\; \Omega,}
\end{eqnarray}
where $T$ is the holomorphic component of the
stress-energy tensor, which is the quasiprimary
field at level $2$ with $a_{2}=-\frac{3}{2}$ in the family
of the identity.
\medskip

I have also checked that (\ref{Lemma21},\,\ref{Lemma2})
reproduces the transformation law of \cite{RS88}
for the quasiprimary field at level $6$ in the
$1$-family.
\bigskip

\noindent {\bf Acknowledgements}
\smallskip

It is a pleasure to thank my PhD supervisor Peter Goddard for much
advice and encouragement.
I also acknowledge useful discussions with A. Kent.

I am grateful to Pembroke College, Cambridge,
for a research studentship.

\end{document}